\documentclass[sigconf, screen]{acmart}
\AtBeginDocument{%
  }

\usepackage[]{collab}
\usepackage{xcolor}
\usepackage{ulem}
\usepackage{pifont}
\usepackage{listings}
\usepackage{xcolor}
\usepackage{caption}
\usepackage{float}
\usepackage{hyperref}
\usepackage[]{collab}
\usepackage{balance}
\usepackage{xcolor}
\usepackage{subfigure}
\usepackage{multirow}
\usepackage{multicol}
\usepackage{tabularx}
\usepackage{siunitx}
\usepackage{float}
\usepackage{color}
\usepackage{diagbox}
\usepackage{caption}
\usepackage{makecell}
\usepackage{colortbl}
\usepackage{tcolorbox}
\usepackage{mdframed}
\usepackage{listings}
\usepackage{booktabs}
\usepackage[utf8]{inputenc}
\usepackage{calligra}
\usepackage[normalem]{ulem}
\usepackage{url}
\usepackage{soul}
\usepackage{nicematrix}
\usepackage{fontawesome5}
\usepackage{textcomp}
\usepackage{stfloats}
\usepackage{verbatim}
\usepackage{graphicx}
\usepackage{amsmath,amsfonts}
\usepackage{algorithmic}
\usepackage{array}
\usepackage[utf8]{inputenc}
\usepackage{enumitem}

\collabAuthor{js}{red}{Junsong}
\collabAuthor{zb}{teal}{Zhuangbin}

\newcommand{\revision}[1]{\textcolor{black}{#1}}

\setcopyright{cc}
\setcctype{by-nc-nd}
\acmDOI{10.1145/3832783.3837420}
\acmYear{2026}
\copyrightyear{2026}
\acmISBN{979-8-4007-2882-2/2026/10}
\acmConference[ASE '26]{Proceedings of the 41st IEEE/ACM International Conference on Automated Software Engineering}{October 12--16, 2026}{Munich, Germany}
\acmBooktitle{Proceedings of the 41st IEEE/ACM International Conference on Automated Software Engineering (ASE '26), October 12--16, 2026, Munich, Germany}
\acmSubmissionID{ase26main-p169-p}
\received{2026-03-26}
\received[accepted]{2026-06-18}

\newcommand\alias{\textsc{AutoSQL}\xspace}

\begin{document}
\title{\alias: Extracting SQL Templates from Imperative ORM Code in Large-Scale Repositories}

\author{Junsong Pu}
\orcid{0009-0002-8309-1384}
\affiliation{%
  \institution{Sun Yat-sen University}
  \city{Zhuhai}
  \country{China}
}
\email{pujs@mail2.sysu.edu.cn}

\author{Yichen Li}
\authornotemark[1]
\orcid{0009-0009-8370-644X}
\affiliation{%
  \institution{Chinese University of Hong Kong}
  \city{Hong Kong}
  \country{Hong Kong}
}
\email{ycli21@cse.cuhk.edu.hk}

\author{Zhuangbin Chen}
\orcid{0000-0001-5158-6716}
\authornote{Corresponding authors.}
\affiliation{%
  \institution{Sun Yat-sen University}
  \city{Zhuhai}
  \country{China}
}
\email{chenzhb36@mail.sysu.edu.cn}

\author{Zhihan Jiang}
\orcid{0009-0003-1988-6219}
\affiliation{%
  \institution{Chinese University of Hong Kong}
  \city{Hong Kong}
  \country{Hong Kong}
}
\email{zhjiang22@cse.cuhk.edu.hk}

\author{Zibin Zheng}
\orcid{0000-0002-7878-4330}
\affiliation{%
  \institution{Sun Yat-sen University}
  \city{Zhuhai}
  \country{China}
}
\email{zhzibin@mail.sysu.edu.cn}






\renewcommand{\shortauthors}{Pu et al.}


\begin{abstract}

Suboptimal SQL queries can significantly degrade the performance of cloud systems, motivating the extraction and auditing of SQL statements before deployment. However, Go ORM frameworks construct SQL imperatively through scattered method-call sequences, making it difficult to statically recover the resulting SQL templates. We present \alias, a system that reconstructs SQL templates from Go ORM code.
\alias constructs a Code Index, a directed graph that captures structural dependencies between functions, types, and global variables as navigable edges. It then traces upstream call chains from ORM invocation sites to identify database-interacting functions as entry points. For each entry point, an LLM agent traverses the Code Index to collect code slices that influence SQL generation, switching to pattern-based search when the graph cannot resolve a retrieval goal. We call this strategy \textit{Hybrid Context Retrieval}. Once sufficient context is collected, the agent synthesizes SQL templates. Evaluation on \revision{a benchmark of 579 test-covered entry points and} 1,186 \revision{runtime-traced} SQL statements \revision{from} five large-scale Go repositories shows that \alias achieves 68.04\% to 72.18\% recall, exceeding the static reachability \revision{baseline} by 11.80\% to 15.94\% and outperforming existing methods by 8.52\% to 21.50\%.

\end{abstract}

\ccsdesc[500]{Software and its engineering~Software maintenance tools}
\ccsdesc[500]{Software and its engineering~Software testing and debugging}

\keywords{SQL Auditing, Object-Relational Mapping, Code Agents, Code Index}

\maketitle

\section{Introduction}

Modern cloud services rely on database systems to provide reliable persistent storage. SQL queries that scan large tables without proper indexing, or that retrieve excessive data through missing filters, execute slowly and occupy database connections for extended periods. When such queries are triggered frequently, connection pools become exhausted and response latency increases for all services sharing the same database~\cite{ORMAntiPattern2014, CloudBugStudy2014}. To prevent inefficient queries from reaching production, developers perform SQL auditing, which reviews the SQL statements an application may execute and checks for performance risks before deployment.

In traditional database programming, developers write SQL as string literals in the source code, such as \texttt{db.Query("DELETE FROM users WHERE tenant\_id = ?", id)}, making every query directly visible for review. However, modern applications increasingly adopt Object-Relational Mapping (ORM) frameworks, which generate SQL from method calls rather than requiring developers to write it directly. ORM frameworks are typically used in two distinct styles. In the \textit{declarative} style, the mapping between code and database schema is fixed through static metadata (e.g., Java annotations like \texttt{@Table(name="users")}), and tools can extract SQL by parsing these declarations. In the \textit{imperative} style, SQL is constructed dynamically. Each method call appends a component to a stateful builder object, and the final SQL emerges from the accumulated state at execution time. For example, a call sequence like \texttt{sess.Where("tenant\_id=?",~id).Delete(\&User\{\})} constructs a \texttt{DELETE} statement through successive state mutations rather than a visible SQL string.

The Java ORM ecosystem predominantly adopts the declarative style, while the Go ORM ecosystem predominantly adopts the imperative style. The declarative style poses little challenge for SQL extraction, as tools can parse static metadata directly. The imperative style, however, creates a fundamental obstacle for SQL auditing, because the SQL statements an application will execute do not exist as inspectable artifacts in the source code. A single SQL statement may depend on code entities scattered across multiple files and modules, including a \texttt{TableName()} method that resolves the target table through runtime configuration, helper functions that progressively append \texttt{WHERE} conditions to a session object, and lifecycle hooks that the framework invokes implicitly to inject additional column assignments. Reconstructing the complete SQL requires tracing all these dependencies and reasoning about how the ORM assembles them at runtime. Our task is to automatically extract such SQL templates from Go source code so that developers can audit them before deployment.

Existing static analysis tools for database code are designed for declarative ORM frameworks and cannot handle stateful, imperative query construction~\cite{Dbridge, SLocator}. Symbolic execution can theoretically explore all execution paths but suffers from path explosion in large codebases~\cite{cusr_sym_survey}. Recent LLM-based code agents offer an alternative through semantic reasoning, but existing code audit agents target security vulnerability detection~\cite{iCodeReviewer, RepoAudit}. Their retrieval strategies are insufficient for the deep cross-boundary context collection that SQL reconstruction demands.

This paper presents \alias, an LLM-based agent that reconstructs SQL templates from Go ORM code. The core challenge is context retrieval, as the agent must locate and collect the scattered code entities that contribute to each SQL statement across the repository. We first construct a Code Index, a directed graph that captures structural dependencies between code entities as navigable edges, and identify entry points for analysis by tracing upstream call chains from ORM invocation sites. For each entry point, the agent collects context through \textit{Hybrid Context Retrieval}, which combines two complementary retrieval modes. The agent traverses the Code Index to efficiently trace the functions, type definitions, and implicit interface relationships that contribute to each SQL statement across files. However, some dependencies require finer-grained reasoning about how values propagate through dynamic language features, which symbol-level structural edges do not capture. For these cases, the agent switches to on-demand pattern-based search, synthesizing targeted regular expressions from its current retrieval goal to locate the relevant code directly in source files. Once sufficient context is collected, the agent synthesizes SQL templates by reasoning about ORM framework behaviors and enumerating control-flow paths that lead to different SQL variants.

We evaluate \alias on five large-scale open-source Go repositories ranging from 38K to 845K lines of code, covering two ORM frameworks and three database dialects. We construct a benchmark of \revision{579 test-covered entry points and} 1,186 \revision{runtime-traced} SQL statements, of which 43.76\% rely on dynamic features that prevent pure static analysis. \alias achieves 68.04\% to 72.18\% recall, surpassing the \revision{DBridge-style~\cite{Dbridge}} static reachability \revision{baseline} by 11.80\% to 15.94\% and outperforming existing code audit agents by 8.52\% to 21.50\%. Ablation results confirm that structural navigation via the Code Index improves recall by 2.35\% to 37.93\% over pattern-based search alone.

This work makes the following major contributions:

\begin{itemize}[noitemsep,leftmargin=5.5mm]
\item We propose \alias, an LLM-based agent that extracts SQL templates from imperative Go ORM code through Hybrid Context Retrieval, combining structural graph navigation with on-demand pattern-based search for implicit dependencies.

\item We construct a benchmark dataset from five Go repositories comprising \revision{579 test-covered entry points and} 1,186 \revision{runtime-traced} SQL statements, of which 43.76\% rely on dynamic features beyond the reach of practical deterministic methods.

\item Evaluation shows that \alias achieves 68.04\% \textasciitilde~72.18\% recall \revision{on the runtime-traced benchmark}, surpassing the \revision{DBridge-style} static reachability \revision{baseline} by 11.80\% \textasciitilde~15.94\% and existing code audit agents by 8.52\% \textasciitilde~21.50\%. Ablation results confirm that both structural graph navigation and pattern-based search contribute to the improvement.
\end{itemize}

\section{Background and Motivation}
\label{sec:background}

\subsection{Object-Relational Mapping Mechanics}

Object-Relational Mapping (ORM) frameworks abstract database operations into application-level method calls, allowing developers to manipulate data without writing SQL directly~\cite{icsoft-ea16,Colley2018-iCCECE}. The framework translates these method calls into SQL at runtime. How this translation is specified determines whether the resulting SQL can be statically extracted from source code. ORM frameworks adopt one of two styles: declarative or imperative.

In the \textit{declarative} style, the mapping between application code and database schema is specified through static metadata fixed at definition time. The Java ORM ecosystem implements this approach through annotations: \texttt{@Table(name="users")} binds a class to the \texttt{users} table, and \texttt{@Column(name="user\_name")} binds a field to a column. Because annotation parameters must be compile-time constants, the schema mapping is statically determined, and tools can extract SQL structures by parsing these declarations without analyzing control flow.

\begin{figure*}
    \includegraphics[width=0.7\textwidth]{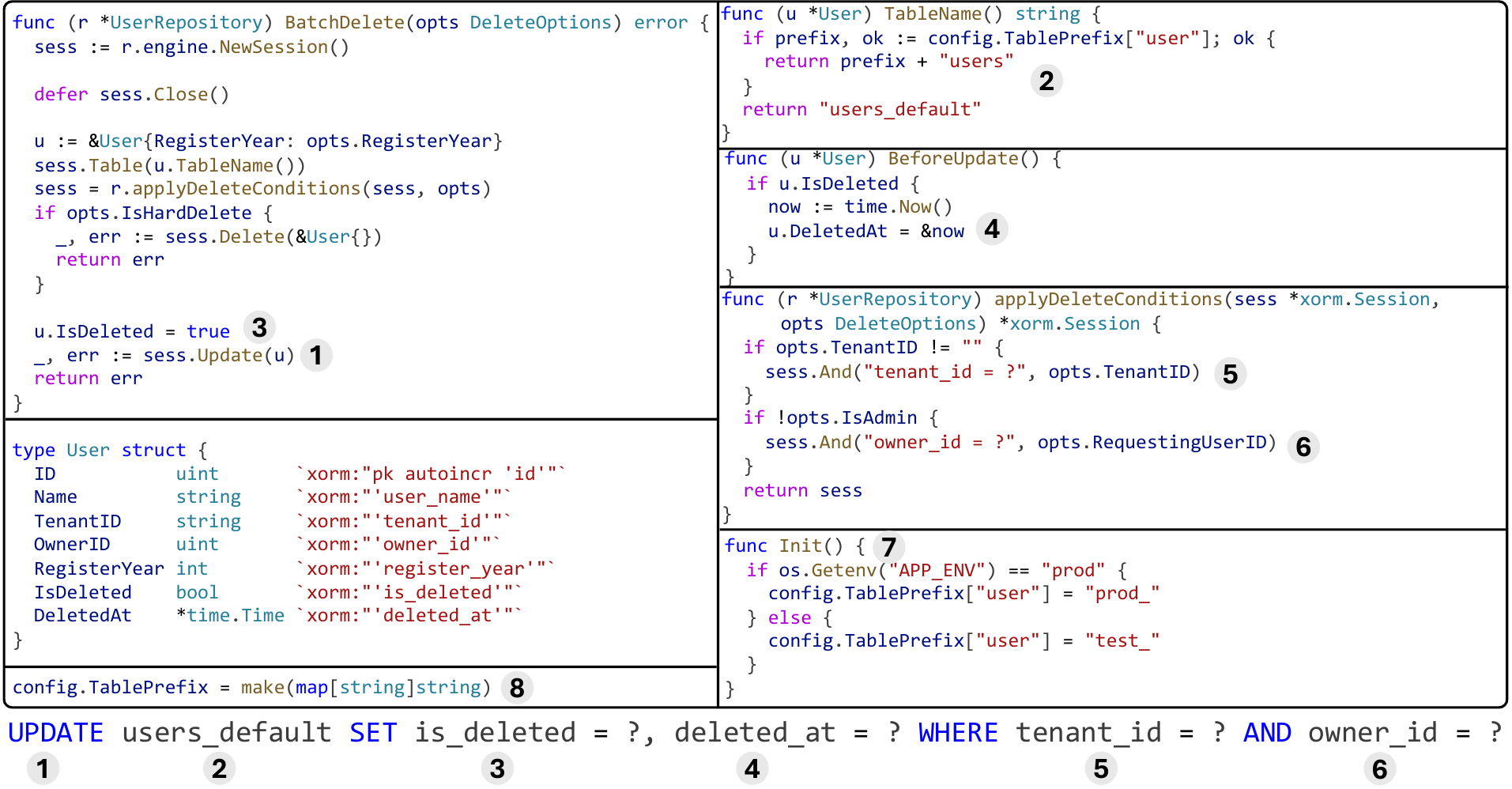}
    \caption{An Example of SQL Generation in XORM}
    \label{fig:orm_example}
\end{figure*}

In the \textit{imperative} style, SQL is constructed dynamically through sequences of method calls at runtime. Each call appends a component, such as a table reference, a filter condition, or a column assignment, to a stateful builder object. The final SQL emerges from the accumulated state when the query is executed. Frameworks such as XORM~\cite{xorm_website} and Beego-ORM~\cite{beego_orm_website} in the Go ecosystem adopt this style. Figure~\ref{fig:orm_example} illustrates an example: when the \texttt{BatchDelete} function executes, the ORM produces the SQL \texttt{UPDATE users\_default SET is\_deleted=?, deleted\_at=? WHERE tenant\_id=? AND owner\_id=?}. We trace how each component of this SQL is assembled from code scattered across the repository.

\subsubsection{Runtime Table Name Resolution}
The table name \texttt{users\_def ault} originates from the \texttt{TableName} method at label \ding{173}. Rather than returning a fixed string, this method contains conditional logic: it checks the global configuration map \texttt{config.TablePrefix} and returns a prefixed or default name accordingly. The configuration map itself is initialized in a separate module at label \ding{178}, based on environment variables. Extracting the table name therefore requires tracing from the ORM invocation to the \texttt{TableName} method definition, and then to the global variable initialization across module boundaries. We refer to this pattern, where schema information is resolved through method dispatch and runtime state, as \textit{runtime schema resolution}.

\subsubsection{Stateful Builder Pattern}
\label{sssec:builder}
The \texttt{WHERE} clause is assembled by the helper function \texttt{applyDeleteConditions}, which appends filter conditions through calls to \texttt{sess.And} at labels \ding{176} and \ding{177}. The session object maintains internal state that accumulates these conditions as it passes through the call chain. When \texttt{Update} is finally called at label \ding{172}, the ORM reads this accumulated state to generate the complete \texttt{WHERE} clause. Extracting these conditions requires following the call chain from \texttt{BatchDelete} into the helper function and tracking how the session state is progressively mutated. This \textit{builder pattern} constructs queries through stateful accumulation across multiple function calls rather than through static SQL strings.

\subsubsection{Lifecycle Hooks}
The \texttt{SET} clause contains two columns: \texttt{is\_de leted}, explicitly assigned at label \ding{174}, and \texttt{deleted\_at}, which appears nowhere in the main function. The second column is injected by the \texttt{BeforeUpdate} hook at label \ding{175}. The ORM framework discovers this hook through Go's structural typing~\cite{FeatherweightGo}: the \texttt{User} struct implicitly satisfies the \texttt{BeforeUpdateProcessor} interface by defining a method with the matching signature, without any explicit \texttt{implements} declaration. The framework detects this relationship at runtime via reflection and invokes the hook automatically before executing the update.

This mechanism, known as a \textit{lifecycle hook}, poses a particular challenge for SQL reconstruction. Because the framework discovers and invokes the hook via reflection, no explicit call edge from the ORM engine to \texttt{BeforeUpdate} appears in the program's call graph. An analysis that follows the call graph will not reach the hook's body, leaving the SQL columns it contributes unrecoverable unless the implicit invocation is explicitly modeled.

\subsection{Motivation}

\subsubsection{Problem Statement}

The three mechanisms above illustrate that a single SQL statement in imperative ORM code is an emergent product of code slices distributed across multiple files and modules. These slices contribute to the final SQL through diverse dependency types, i.e., method calls, global variable reads, and implicit interface implementations, and their roles only become apparent when reasoning about how the ORM assembles them at runtime. The resulting SQL does not exist as an inspectable artifact in the source code, preventing developers from auditing queries before deployment. Our task is to automatically reconstruct SQL templates from such code: given a Go repository that uses an ORM framework, extract the SQL statements the application may execute.

\subsubsection{Challenges of Static Approaches}
\label{sssec:existing_limitation}

Existing static analysis tools for database code are designed for declarative ORM frameworks. Tools such as DBridge~\cite{Dbridge} and SLocator~\cite{SLocator} extract SQL by parsing metadata annotations (\textit{e.g.}, \texttt{@Entity}, \texttt{@NamedQuery}) to build static models of query logic. As DBridge itself notes, these techniques cannot handle stateful, imperative query construction where the SQL structure is an emergent property of builder state accumulation rather than a static definition.

Existing Go-native tools serve different purposes. Vulnerability scanners such as \texttt{govulncheck}~\cite{go_vulncheck} detect known CVEs in dependencies, and SQL linters such as \texttt{sqlcheck}~\cite{sqlcheck_linter} check SQL strings for anti-patterns but require the SQL to already exist as input. Neither attempts to reconstruct SQL from ORM code. More general program analysis techniques such as symbolic execution could theoretically enumerate all SQL variants by exploring execution paths, but suffer from path explosion in real-world codebases~\cite{cusr_sym_survey}.

\subsubsection{LLM Code Agents as an Alternative}

Large Language Model (LLM) code agents offer a different approach: reasoning about code semantics rather than matching pre-defined syntactic patterns. This distinction is critical for imperative ORM code, where the same database operation can be expressed through diverse code structures. Consider the \texttt{BeforeUpdate} hook in Figure~\ref{fig:orm_example}, which sets the timestamp at label \ding{175}. The same soft-delete logic could equally be implemented by manipulating low-level clause builders, using variable assignments, or concatenating raw SQL strings. A rule-based analyzer would require explicit patterns for each variant, whereas an LLM can infer that these syntactically different forms produce equivalent effects and synthesize the corresponding SQL. Recent work has demonstrated LLM agents' effectiveness in repository-level code understanding tasks including code auditing~\cite{iCodeReviewer, RepoAudit}, automated program repair~\cite{yang2024sweagent, xia2025agentless}, type inference~\cite{TypeGen2023}, and root cause analysis~\cite{COCA2025}.

However, existing code audit agents are primarily designed for security vulnerability detection rather than semantic reconstruction of database operations. Extracting SQL templates from Go ORM code presents unique challenges: tracking the state accumulation of builder objects across multiple function calls, understanding implicit hook mechanisms such as \texttt{BeforeUpdate} that transform the semantics of an operation, and reconstructing complete SQL statements rather than merely detecting vulnerability patterns. This demands deep cross-boundary context retrieval that current agent architectures do not adequately support.

\section{Methodology}

\subsection{Overview}

\begin{figure}
\includegraphics[width=1\columnwidth]{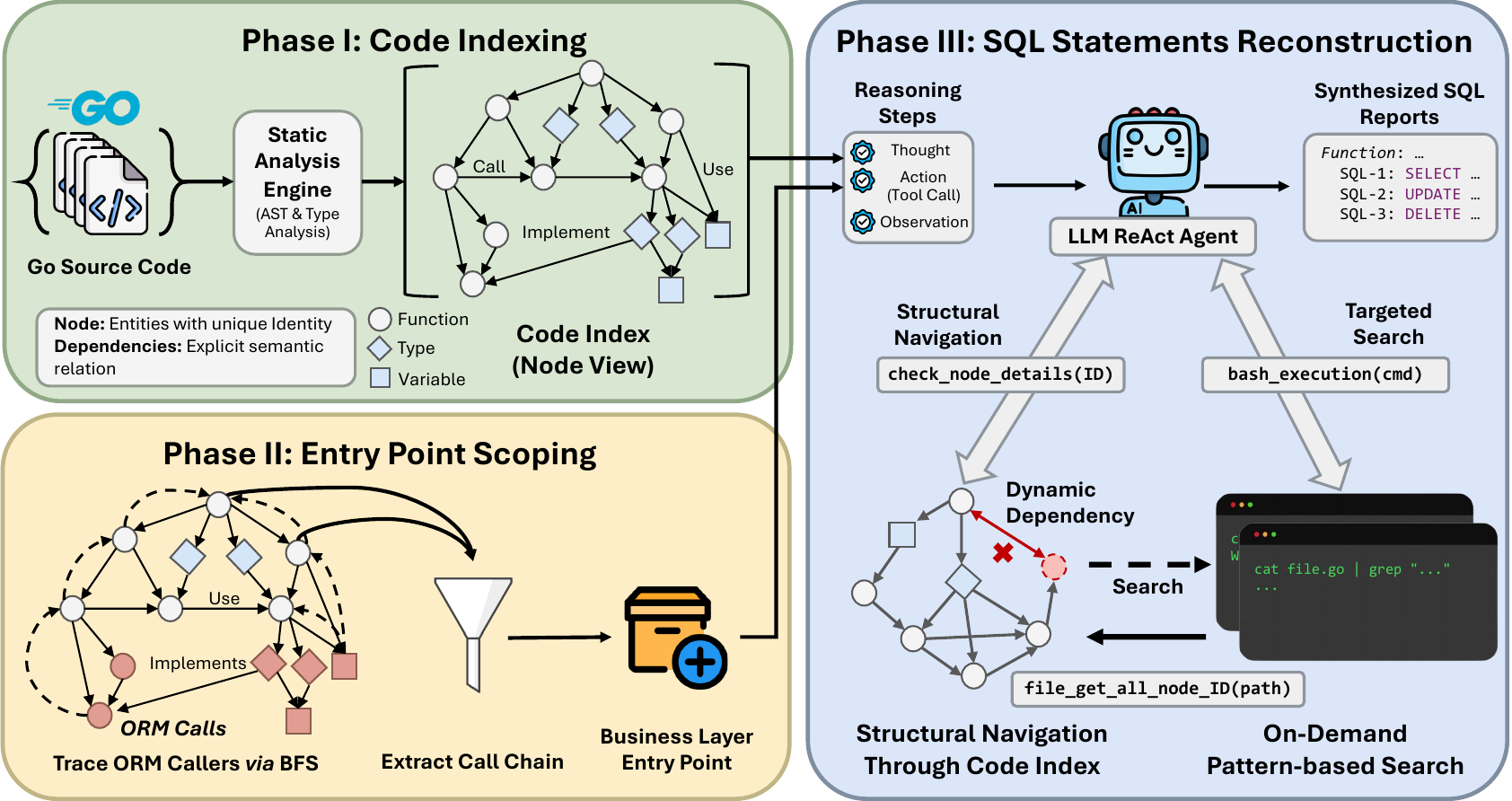}
\vspace{-6pt}
\caption{Overview of \alias Methodology}
\vspace{-6pt}
\label{fig:main}
\end{figure}

As shown in Figure~\ref{fig:main}, \alias reconstructs SQL templates from Go ORM code through three phases. In Phase~I, we construct a Code Index, a directed graph that captures structural dependencies between code entities as a navigable backbone for the repository. In Phase~II, we identify all direct ORM invocation sites and trace their upstream call chains within the Code Index, narrowing the analysis scope to functions that interact with the database. In Phase~III, an LLM Agent collects code context relevant to each entry point through \textit{Hybrid Context Retrieval}, which combines structural navigation over the Code Index with on-demand pattern-based search for dependencies beyond graph coverage, and then synthesizes SQL templates from the collected context. \revision{The Code Index and Hybrid Context Retrieval are framework-agnostic; ORM-specific knowledge (terminal APIs, struct tag semantics, hook conventions) is encapsulated in the system prompt and can be adapted to new imperative ORM frameworks independently.}

\subsection{Phase I: Code Indexing}

We define the Code Index as a directed graph $G = (N, E)$ for navigation, where nodes are code entities and edges are dependency relations. We call these relations structural dependencies because they are explicit in program syntax and connect symbols through declaration and reference.

\subsubsection{Indexing Strategy}

The Code Index is designed to slice the entire repository at the symbol level and capture the structural dependencies between these symbols. While SQL reconstruction inherently requires reasoning about data flow, performing full-program data-flow analysis is practically infeasible for large-scale repositories. Conversely, limiting the scope to intra-procedural statement-level analysis offers little value, as dynamic SQL construction typically spans multiple functions and files.

Therefore, rather than building complex statement-level dependence graphs, we keep the graph lightweight to serve purely as a navigation backbone and delegate the actual value-flow reasoning to the LLM. In practice, the scattered slices needed for SQL reconstruction, such as struct definitions and helper functions, are highly reachable via structural relations. By storing the complete source code at each symbol node, we provide sufficient context for the LLM to deduce intra-procedural data flow directly from the retrieved code slices. This design choice is supported by prior findings~\cite{li2024enhancing}, which demonstrate that LLMs reason more effectively over raw source code than over parsed intermediate representations.

\subsubsection{Graph Structure}

The graph contains three categories of nodes and four categories of edges, summarized in Tables~\ref{tab:nodes} and~\ref{tab:edges}. Each node stores its source code and is uniquely indexed by a \textit{Global Identity} tuple (module, package, and symbol name) for precise retrieval. The node's metadata includes its file location, function signature, and field types. Crucially, every symbol reference within the metadata is resolved to a Global Identity to act as a direct link. For example, the metadata of \texttt{BatchDelete} (Figure~\ref{fig:orm_example}) links \texttt{UserRepository} and \texttt{DeleteOptions} via their Global Identities, making these dependencies immediately reachable for the LLM Agent.
With the Code Index constructed, the next step is to identify which functions warrant analysis.

\begin{table}[t]
\centering
\caption{Node Categories in the Code Index}
\label{tab:nodes}
\small
\begin{tabular}{p{1.8cm}p{5.8cm}}
\toprule
\textbf{Node Type} & \textbf{Description} \\
\midrule
\textit{Function} & All function and method definitions in the repository, including both standalone functions and methods attached to types. \\
\midrule
\textit{Type} & Struct and interface definitions. For structs, the node records field types and nested substruct definitions. For interfaces, it stores method signatures. \\
\midrule
\textit{Variable} & Global variable declarations along with their type information. \\
\bottomrule
\end{tabular}
\end{table}

\begin{table}[t]
\centering
\caption{Edge Categories in the Code Index}
\label{tab:edges}
\small
\begin{tabular}{p{1.3cm}p{6.4cm}}
\toprule
\textbf{Edge Type} & \textbf{Description} \\
\midrule
\textit{Call} & Direct invocation relationships between functions, including both regular function calls and method calls. \\
\midrule
\textit{Use} & References from a function to other symbols, such as reading or writing a global variable, creating a local variable of a certain type, or assigning a function to a variable. \\
\midrule
\textit{Implement} & A type implicitly implements an interface by providing matching methods, without explicit declaration. \\
\midrule
\textit{Inherit} & One type directly contains another inside its definition, gaining access to the embedded type's fields and methods. \\
\bottomrule
\end{tabular}
\end{table}

\subsection{Phase II: Entry Point Scoping}

Directly analyzing every function within a repository-scale codebase is computationally wasteful, as the vast majority of functions do not interact with the database. To focus the Agent's reasoning power on database interaction logic, we employ a scoping strategy that identifies all direct ORM invocation sites and traces their upstream call chains within the Code Index.

\subsubsection{Identification of ORM Invocation Sites}
The process begins by identifying functions that directly invoke ORM framework methods that execute SQL statements against the database, such as \texttt{Insert}, \texttt{Delete}, \texttt{Update}, \texttt{Get}, and \texttt{Find} on database session objects. Specifically, we rely on a predefined list of these terminal method signatures tailored to the supported ORM frameworks. These terminal methods trigger actual database operations. Functions containing these terminal method calls serve as the initial set of candidates for entry point analysis.

\subsubsection{Upstream Call Chain Tracing}
However, analyzing these invocation sites in isolation may sometimes be insufficient. In well-structured Go applications, variables that influence query conditions are frequently constructed by upstream callers and passed down as parameters. For example, in Figure~\ref{fig:orm_example}, the terminal \texttt{Update} method is invoked directly within \texttt{BatchDelete} (Label \ding{172}). While the local control flow captures the core database operation, the specific query constraints can be heavily influenced by the \texttt{opts} parameter. To better understand how these fields might be populated in practical scenarios, it is beneficial to trace the call chain upstream to identify the callers of \texttt{BatchDelete}. Relying solely on the execution site without its upstream context might yield overly generic or incomplete SQL templates.

To resolve this, we trace the inverse call edges from ORM invocation sites up to a bounded depth $D$. Formally, let $\mathcal{S}_{orm} \subset N$ be the set of functions that directly invoke ORM APIs, and define the entry point set as $\mathcal{E} = \bigcup_{k=0}^{D} \mathcal{C}_k$, where $\mathcal{C}_k = \{ u \in N \mid \exists v \in \mathcal{C}_{k-1}, (u, v) \in E_{call} \}$ and $\mathcal{C}_0 = \mathcal{S}_{orm}$. Here $E_{call}$ denotes the set of \textit{Call} edges in the Code Index. For each entry point, Phase II records the call chain to the ORM invocation site for subsequent analysis.

\subsection{Phase III: SQL Statement Reconstruction}

For each entry point identified in Phase II, the Agent reconstructs SQL statements through iterative context collection followed by single-turn SQL synthesis. Guided by predefined ORM framework knowledge, the Agent operates under the ReAct (Reasoning and Acting) paradigm~\cite{yao2022react}, issuing tool invocations as \textit{Actions}, receiving code slices and dependency information as \textit{Observations}, and reasoning about data flow in \textit{Thought} steps. This iterative process continues until the Agent accumulates sufficient context to synthesize SQL templates. We describe the retrieval toolset, the context collection strategy, and the SQL synthesis procedure below.

\subsubsection{Retrieval Toolset}

The Agent accesses the Code Index and source files through three tools, summarized in Table~\ref{tab:tools}. These tools enable both structural navigation through the Code Index and pattern-based search through source files, with mechanisms to bridge between the two retrieval modes.

\begin{table}[h]
\centering
\caption{Retrieval Tools for Code Navigation}
\label{tab:tools}
\small
\begin{tabular}{p{2cm}p{5.5cm}}
\toprule
\textbf{Tool} & \textbf{Description} \\
\midrule
\texttt{check\_node\_ details} & Accepts a Global Identity to query the Code Index, returning the node's source code, its metadata, and its forward and backward structural edges to other entities in the graph. It prompts if the node was already queried in the same session.
 \\
\midrule
\texttt{file\_get\_all\_ node\_identity} & Accepts a file path and extracts Global Identities of all entities defined within it. \\
\midrule
\texttt{bash} & Executes shell commands such as \texttt{grep} to read source files, perform pattern-based search, and access configuration files in the repository. \\
\bottomrule
\end{tabular}
\end{table}

\subsubsection{Context Collection}

Given an entry point from Phase II, the Agent receives the call chain to the ORM invocation site along with the source code of functions in that chain. The Agent reasons about data flow in \textit{Thought} steps, understanding how variables are transformed and how query conditions accumulate across function calls. The Agent then collects additional code context beyond the call chain, such as type definitions, lifecycle hooks, and configuration values that influence the generated SQL.

To collect these dependencies, the Agent uses a strategy called \textit{Hybrid Context Retrieval}, which combines structural navigation through the Code Index with on-demand pattern-based search. The Agent primarily navigates the Code Index, which records structural relationships between functions, types, and variables as graph edges. This allows the Agent to trace dependencies across files by following edges. For example, when following the call to \texttt{TableName} in Figure~\ref{fig:orm_example}, the Code Index directs the Agent to the \texttt{User} struct's \texttt{TableName} method at Label \ding{173}, rather than all methods named \texttt{TableName} across the repository.

However, structural navigation alone cannot satisfy all retrieval goals. The system prompt instructs the Agent to switch to pattern-based search when the Code Index cannot resolve the current retrieval goal to a specific code location. This occurs under two conditions. First, the Agent switches when no outgoing edge from the current node leads toward the retrieval goal, indicating that the target lies outside the graph's scope, such as configuration files or dynamically constructed calls. Second, the Agent switches when the retrieval goal requires information that graph edges do not encode. The symbol-level graph records which functions access a given symbol, but does not record details such as whether each access is a read or write, which arguments are passed, or which map key is involved. When the retrieval goal requires filtering based on such details, the Agent synthesizes a search pattern that encodes the precise syntactic signature of the target, then executes pattern-based search to locate it directly.

Pattern-based search runs on demand by turning the retrieval goal into a concrete search pattern that matches the target code form. Consider locating where \texttt{TablePrefix} is initialized in Figure~\ref{fig:orm_example}: the \texttt{TableName} method at label \ding{173} reads \texttt{TablePrefix["user "]}, and the Code Index shows all functions that access this variable. The Agent understands the semantic goal is to find where the map key \texttt{"user"} is assigned a value, not just any access to \texttt{TablePrefix}. It constructs the pattern \texttt{TablePrefix[}\texttt{\textbackslash"user\textbackslash"]\textbackslash s*=} that specifically matches assignment to this key, locating the initialization at label \ding{178}. 

This design avoids a common tradeoff in existing approaches. Fixed pattern matching needs a predefined library of patterns for many code structures, which does not transfer well across projects with different coding styles. Whole program data flow can track how values propagate, but it needs pointer analysis to model aliasing and does not scale to large codebases. The Agent instead builds the search pattern from the current semantic goal and avoids both a fixed rule set and global analysis.

Once the Agent discovers relevant code through pattern-based search, it uses the \texttt{file\_get\_all\_node\_identity} tool to map the discovery back to the Code Index, enabling further structural navigation from that point.This bidirectional switching provides two advantages. First, it allows the Agent to maintain high precision for dependencies with clear structural paths. Second, it effectively handles edge cases where the Code Index lacks the necessary semantic information.

The Agent stops context collection when \revision{neither structural traversal nor pattern-based search yields new SQL-relevant code slices along the current call chain}. \revision{The} context window \revision{limit of} 90\% of its maximum \revision{serves as a hard cutoff to prevent context overflow}.

\subsubsection{SQL Template Synthesis}

Once the Agent accumulates sufficient context, it synthesizes SQL templates through a single turn Chain of Thought inference~\cite{wei2022chain}. The system prompt guides the Agent to perform the following reasoning steps.

\textbf{ORM Implicit Behavior Analysis.} The Agent identifies implicit ORM semantics that alter the generated SQL without explicit invocation in the application code. Guided by framework-specific knowledge in the system prompt, the Agent analyzes data models for two primary implicit behaviors: struct tags and lifecycle hooks. For tag semantics, an annotation like \texttt{xorm:"deleted"} dictates that the framework internally transforms \texttt{DELETE} operations into \texttt{UPDATE} statements. For lifecycle hooks, if a struct implements a specific interface method like \texttt{BeforeUpdate}, the Agent recognizes that the framework automatically triggers this hook to inject or modify column values (e.g., updating a timestamp) right before query execution, ensuring the final SQL template captures these hidden state mutations.

\textbf{Control-Flow Enumeration.} \revision{This is a prompt-level design: we instruct the Agent to enumerate possible control-flow paths leading to distinct SQL variants before synthesizing each template, allowing the model to prune infeasible paths through semantic reasoning.} We instruct the Agent to identify all SQL generation points, which are locations where ORM APIs are invoked to execute SQL statements, within the collected code and enumerate control-flow paths that lead to different SQL variants. Conditional branches that affect table names, column selections, or \texttt{WHERE} conditions result in separate SQL templates. Table names are resolved from \texttt{TableName()} methods or derived from struct names. The placeholder \texttt{<*>} is used only when an attribute cannot be inferred from the collected code snippets. Finally, the Agent synthesizes SQL statements according to the user-specified database dialect.

\subsubsection{Example Walkthrough}

To illustrate the complete reconstruction process, we use the example from Figure~\ref{fig:orm_example}, assuming an upstream tracing depth of $D=0$ in Phase II.

\textbf{Context Collection.} Starting from \texttt{BatchDelete}, the Agent traces forward to identify the SQL execution point at \texttt{sess.Update} (Label \ding{172}), and the explicitly set soft-delete flag (Label \ding{174}). Tracing backward, it identifies the \texttt{BeforeUpdate} hook via inspecting its type metadata (Label \ding{175}), and follows the call graph to \texttt{applyDelete
Conditions} (Labels \ding{176}, \ding{177}) to gather \texttt{WHERE} conditions. To resolve the target table, it visits \texttt{TableName} (Label \ding{173}), where it encounters the global \texttt{TablePrefix} (Label \ding{179}). Requiring dataflow context beyond the graph, the Agent switches to pattern-based search to locate the variable's initialization in \texttt{Init} (Label \ding{178}).

\textbf{SQL Synthesis.} From the collected context, the Agent deduces that the explicitly set \texttt{IsDeleted} flag (Label \ding{174}) triggers the \texttt{Before Update} hook (Label \ding{175}) to populate the \texttt{DeletedAt} timestamp. Applying ORM implicit behavior analysis, it maps these fields to the \texttt{is\_deleted} and \texttt{deleted\_at} columns. The Agent also integrates the \texttt{tenant\_id} and \texttt{owner\_id} conditions gathered from \texttt{applyDele teConditions} (Labels \ding{176}, \ding{177}). To simplify the demonstration, we assume the logic at both labels is triggered. Finally, by enumerating control-flow paths in \texttt{TableName} (Label \ding{173}), the Agent generates three complete SQL templates: \texttt{UPDATE \{prod, test\}\_users SET is\_deleted=?, deleted\_at=? WHERE tenant\_id=? AND owner\_id=?} and \texttt{UPDATE users\_ default SET is\_deleted=?, deleted\_at=? WHERE tenant\_id=? AND owner\_id=?}.
\section{Study Design}

To evaluate the effectiveness and efficiency of \alias for SQL auditing in large-scale Go applications, we conducted a comprehensive empirical study to answer the following research questions:

\begin{itemize}[noitemsep,leftmargin=5.5mm]
    \item \textbf{RQ1 (Effectiveness):} How does \alias compare to existing approaches in recall and accuracy?

    \item \textbf{RQ2 (Efficiency):} How does \alias compare to existing approaches in token and time cost?

    \item \textbf{RQ3 (Ablation Study):} \revision{How does \alias's Hybrid Context Retrieval compare against alternative retrieval strategies?}

    \item \textbf{RQ4 (Parameter Sensitivity):} How does the upstream tracing depth $D$ affect recall and cost?
\end{itemize}

\subsection{Subject Systems}
We selected five large-scale, open-source Go repositories to serve as our experimental dataset. To ensure experimental representativeness, our selection prioritizes projects characterized by widespread industrial adoption, substantial codebase scale, and extensive reliance on ORM frameworks involving complex dynamic SQL construction. All methods in our evaluation share the same Code Index built by \textsc{Abcoder}~\cite{abcoder} as a one-time offline preprocessing step. Table~\ref{tab:subjects} summarizes the statistics of the subject systems, including the indexing time for each repository.

\begin{table*}[t]
\centering
\caption{Statistics of Subject Systems. LoC (Lines of Code) excludes comments and blank lines. Index Time is the wall-clock time for building the Code Index.}
\label{tab:subjects}
\small{%
\begin{tabular}{l l r r r l l r r r}
\toprule
\textbf{Repository} & \textbf{Description} & \textbf{Stars} & \textbf{Go Files} & \textbf{LoC} & \textbf{ORM} & \textbf{Dialect} & \textbf{SQLs} & \textbf{Test Cases} & \textbf{Index (s)} \\
\midrule
\textbf{Answer~\cite{apache-answer}} & Q\&A platform for teams & 15.3k & 434 & 55,961 & XORM & SQLite & 145 & 70 & 32 \\
\textbf{Gitea~\cite{gitea}} & Self-hosted Git service & 52.9k & 2,789 & 332,970 & XORM & SQLite & 454 & 272 & 172 \\
\textbf{Go-Admin~\cite{go-admin}} & Admin panel framework & 8.9k & 291 & 37,939 & Custom & MySQL & 85 & 46 & 14 \\
\textbf{Grafana~\cite{grafana}} & Analytics \& monitoring platform & 71.5k & 5,085 & 844,968 & XORM & MySQL & 219 & 95 & 1,221 \\
\textbf{Harbor~\cite{harbor}} & Cloud-native registry & 27.3k & 1,562 & 161,870 & Beego & PostgreSQL & 283 & 96 & 173 \\
\midrule
\textbf{Total} & & & 10,161 & 1,433,708 & & & 1,186 & 579 & 1,612 \\
\bottomrule
\end{tabular}%
}
\end{table*}

\subsection{Dataset Construction}
\label{sec:ground_truth}

We constructed the evaluation dataset through two stages: \textit{runtime collection} and \textit{test case sampling}.

\textbf{Runtime Collection.}
First, to capture realistic data, we instrumented the source code of each subject system by modifying the underlying ORM configurations. We injected logging hooks to intercept database operations, recording both the generated SQL statements and their triggering call stacks during the execution of standard integration and regression test suites.

\textbf{Test Case Sampling.}
To establish the ground truth, we employed random sampling to select a subset of test cases from these test suites. Three domain experts then analyzed the corresponding SQL statements and execution paths captured from the sampled tests. Ultimately, we selected 579 integration test cases from these repositories, collecting 1,186 SQL statements. To ensure a fair comparison against the collected ground truth, we restrict all auditing methods to analyze only the entry points whose call chains intersect with the ground truth call stacks.

\revision{\textbf{Evaluation Scope.}
Our ground truth is derived from runtime SQL logs captured during official integration and regression test execution. Recall is measured only against these runtime-traced SQL statements, and all methods are compared on the same set of call-stack-intersecting entry points. Entry points not exercised by any test lack independent ground truth and are therefore excluded from quantitative evaluation, as manual labeling is subjective and does not scale, while writing code to trigger uncovered paths amounts to authoring new integration tests rather than an independent evaluation method. The benchmark thus evaluates the task of recovering SQL templates for test-covered database-interacting entry points. AutoSQL's pipeline is applicable to all candidate entry points identified in Phase~II (2,332 in total across the five repositories), but quantified recall and error analysis are scoped to this runtime-traced benchmark.}

\subsection{Comparative Methods}

\hspace{\parindent} To answer \textbf{RQ1} and \textbf{RQ2}, we evaluate \alias against two representative agent-based code auditing frameworks and a \textit{Static Reachability} baseline \revision{derived from the capability boundaries of DBridge-style~\cite{Dbridge} deterministic approaches}.

\begin{itemize}[noitemsep,leftmargin=5.5mm]
    \item \textbf{\textsc{AutoSQL}:} Our proposed approach using the Hybrid Context Retrieval strategy. We set the upstream tracing depth $D=4$ for entry point scoping. To ensure a fair comparison to baseline methods, we did not inject the ORM call chain.

    \item \textbf{Static Reachability:} This baseline represents the \revision{practical capability boundary of DBridge-style~\cite{Dbridge} deterministic approaches on our dataset}. During the labeling process, we inspected each SQL construction path and marked it as \textit{statically reachable} if the code contains none of the following three dynamic features: stateful builder patterns, reflection-based hook invocation, or runtime schema resolution. The recall of this baseline reflects the proportion of SQL statements that can be resolved through \revision{such} deterministic approaches.

    \item \textbf{iCodeReviewer:} A general-purpose code audit agent that uses an expert system to synthesize prompts based on predefined rules~\cite{iCodeReviewer}. We configure the rules for SQL reconstruction as follows: given a target function containing ORM execution APIs, the system extracts its direct callers, referenced types, and global variables to construct a single prompt. The agent then performs one-shot LLM inference to generate the audit result without iterative refinement.

    \item \textbf{RepoAudit:} A general-purpose code audit agent that navigates the codebase solely through a structural index without shell commands~\cite{RepoAudit}. Since the original RepoAudit only explores the codebase along call relationships, we re-implemented its index with type reference edges to support SQL reconstruction. It employs a ReAct-style reasoning loop, iteratively collecting function dependencies across multiple turns before synthesizing the final audit output.
\end{itemize} 

\revision{To answer \textbf{RQ3}, we compare \alias against three baselines that represent the mainstream retrieval and control paradigms employed by general-purpose coding agents. A recent empirical observation supports this comparison strategy: mini-SWE-agent~\cite{mini-swe-agent}, a minimal ($\sim$100-line) bash-only ReAct scaffold, achieves 76.8\% on SWE-bench Verified~\cite{swebench_website} with Claude Opus 4.5, only 2.4 percentage points below the top-performing Live-SWE-agent~\cite{xia2025livesweagentsoftwareengineeringagents} (79.2\%) with the same model. SWE-bench Verified is a human-curated subset of 500 software engineering tasks in which annotators confirmed that each problem description is unambiguous, the test patch is correct, and the task is solvable from the available context; in other words, each instance constitutes a well-defined task. That a minimal scaffold achieves comparable performance to the top agent on these well-defined tasks suggests that scaffold complexity alone may matter less than the choice of retrieval paradigm when the underlying LLM is strong. Motivated by this hypothesis, we identify three widely-used paradigms in the literature: (a) shell/search-based ReAct, as exemplified by SWE-Agent~\cite{yang2024sweagent}; (b) embedding-based RAG retrieval~\cite{Wang2025CodeRAGBench}; and (c) hierarchical multi-agent decomposition~\cite{hong24metagpt}. Proprietary coding agents (e.g., Codex~\cite{openai_codex_agent}) are not used as baselines because their internal retrieval, prompting, and control policies are not transparent or independently configurable, precluding controlled comparison. Moreover, much of their engineering targets open-ended tasks where requirements are underspecified and iterative clarification and planning are central. As the mini-SWE-agent observation above suggests, such additional scaffold complexity may not be the dominant performance factor for well-defined tasks. AutoSQL's task boundary is well-defined: given a database-interacting entry point, recover its SQL templates. Comparing retrieval paradigms under the same LLM therefore isolates the relevant variables more rigorously. Our three baselines directly instantiate these paradigms, collectively covering the core retrieval and control mechanisms of general-purpose coding agents.}

\begin{itemize}[noitemsep,leftmargin=5.5mm]
    \item \textbf{Bash-Only Agent:} A standard ReAct agent equipped solely with shell tools. This baseline represents the raw capability of an LLM to navigate the codebase using unstructured pattern-based search without any structural guidance. \revision{It is architecturally equivalent to mini-SWE-agent~\cite{mini-swe-agent}, making it a competitive, non-trivial baseline.}

    \item \textbf{Vector-RAG Agent:} An agent equipped with a vector search tool in addition to standard shell utilities. The codebase is segmented into 30-line chunks and indexed using the \texttt{doubao-emb edding-large\_text-250515} model. The agent invokes this tool with keywords to retrieve relevant code snippets based on vector similarity.

    \item \textbf{Multi-Agent System:} A hierarchical framework where a \textit{Manager} agent decomposes the auditing task and delegates specific sub-tasks, such as function analysis and symbol lookup, to subordinate \textit{Worker} agents. All agents in this system are equipped with shell tools to navigate the codebase.
\end{itemize}

To answer \textbf{RQ4}, we evaluate \alias with different upstream tracing depth values $D \in \{0, 2, 4\}$ to analyze the impact of this parameter on recall and token consumption.

\subsection{Experimental Configuration}

\hspace{\parindent}
\textbf{Models.} For \textbf{RQ1} \textasciitilde~\textbf{RQ3}, we evaluate the baseline methods using Kimi K2 \cite{moonshot2025Kimi} and Claude Sonnet 4.5 \cite{anthropic2025claude}, both at a temperature of 0.1. In contrast, iCodeReviewer is evaluated using Kimi K3 and Claude Sonnet 4.6. For \textbf{RQ4}, the baseline methods are evaluated exclusively on Kimi K2.

\textbf{Metrics.} We employ four key metrics for evaluation. All metrics involving correctness judgments use LLM-assisted evaluation followed by domain expert review to ensure accuracy.

\begin{itemize}[noitemsep,leftmargin=5.5mm]
    \item \textbf{Recall:} The ratio of ground-truth SQL statements matched by a prediction from a corresponding entry point. A match requires the same operation and primary table, compatible core query semantics, and coverage of all required fields. Additional fields are allowed.

    \item \revision{\textbf{Estimated Precision (P):} Since the runtime-traced ground truth does not cover all SQL templates the code may produce, a predicted SQL absent from the logs is not necessarily a false positive. We estimate precision through sampling: for each method/repository pair, we sample 20 unmatched predictions (1,200 samples in total) and label each as \textit{false positive} (strictly infeasible given the code) or \textit{possible positive} (supported by the code but absent from runtime logs). The estimated precision is $P = (\mathit{predicted} - \mathit{unmatched} \times \mathit{sampled\_fpr}) / \mathit{predicted}$, where $\mathit{sampled\_fpr}$ is the sampled false-positive rate. Here, $\mathit{unmatched}=\mathit{predicted}-\mathit{hits}$, with $\mathit{hits}$ counting distinct predictions matched to ground-truth SQLs.}
    
    \item \textbf{Token Cost:} This tracks the average number of input and output tokens consumed per entry point analyzed. It serves as a direct proxy for economic cost and inference latency. Average token usage is weighted by the number of test cases across all repositories.

    \item \textbf{Error Categories:} For each ground truth SQL that has no matching prediction, we classify the failure mode into three categories: (1) \textit{Wildcard Missing}, where the predicted SQL uses a wildcard for an attribute whose value is fixed by code and not determined by runtime inputs; (2) \textit{Condition Missing}, where conditions like \texttt{WHERE} clauses or \texttt{JOIN} conditions are incomplete or inconsistent; and (3) \textit{Not Predicted}, where the method produces no SQL that matches that ground truth statement.
\end{itemize}

\textbf{Execution Environment.} We conducted all auditing experiments on a Linux server running Debian 10. The system operates within a KVM virtualized environment configured with 32 virtual CPUs and 64GB of RAM. We accessed the Kimi K2-0905 model via an internally deployed service and the Claude Sonnet 4.5 model via the Anthropic API.

\section{Experiment Results}\label{sec:results}

\begin{table*}[t]
\centering
\caption{\textbf{RQ1: Effectiveness Analysis.} Recall (\%) and \revision{Estimated Precision (P, \%)}. Processing time (T, in minutes, excluding code index construction) is reported for Kimi only. iCR = iCodeReviewer, RA = RepoAudit, AS = \alias}
\label{tab:rq1}

\small
\vspace{0.3em}

\resizebox{\textwidth}{!}{%
\begin{tabular}{l r c c c c c c c c c c c c c c c c c}
\toprule
\multirow{3}{*}{\textbf{Repo}} & \multirow{3}{*}{\textbf{GT}} & \multirow{3}{*}{\textbf{Static Reachability}} & \multicolumn{6}{c}{\textbf{Kimi}} & \multicolumn{6}{c}{\textbf{Claude Sonnet}} & \multicolumn{3}{c}{\textbf{Time (min)}} \\
\cmidrule(lr){4-9} \cmidrule(lr){10-15} \cmidrule(l){16-18}
& & & \multicolumn{2}{c}{\textbf{iCR}} & \multicolumn{2}{c}{\textbf{RA}} & \multicolumn{2}{c}{\textbf{AS}} & \multicolumn{2}{c}{\textbf{iCR}} & \multicolumn{2}{c}{\textbf{RA}} & \multicolumn{2}{c}{\textbf{AS}} & \multirow{2}{*}{\textbf{iCR}} & \multirow{2}{*}{\textbf{RA}} & \multirow{2}{*}{\textbf{AS}} \\
\cmidrule(lr){4-5} \cmidrule(lr){6-7} \cmidrule(lr){8-9} \cmidrule(lr){10-11} \cmidrule(lr){12-13} \cmidrule(lr){14-15}
& & & R & \revision{P} & R & \revision{P} & R & \revision{P} & R & \revision{P} & R & \revision{P} & R & \revision{P} & & & \\
\midrule
\textbf{Answer} & 145 & 54.48 & 33.10 & \revision{36.03} & 25.52 & \revision{81.21} & \textbf{61.38} & \revision{\textbf{85.99}} & 46.21 & \revision{43.10} & 61.38 & \revision{73.80} & \textbf{71.72} & \revision{\textbf{82.42}} & 1.8 & 15.0 & 29.7 \\
\textbf{Gitea} & 454 & 48.90 & 68.06 & \revision{48.88} & 56.83 & \revision{\textbf{93.35}} & \textbf{71.37} & \revision{91.65} & 58.59 & \revision{45.67} & 71.15 & \revision{79.03} & \textbf{74.89} & \revision{\textbf{93.96}} & 6.6 & 54.5 & 81.9 \\
\textbf{Go-Admin} & 85 & 30.59 & 42.35 & \revision{61.00} & 36.47 & \revision{85.95} & \textbf{75.29} & \revision{\textbf{86.58}} & 41.18 & \revision{62.87} & 69.41 & \revision{93.12} & \textbf{71.76} & \revision{\textbf{94.85}} & 0.7 & 15.2 & 24.0 \\
\textbf{Grafana} & 219 & 85.84 & 71.69 & \revision{69.14} & 68.49 & \revision{89.27} & \textbf{84.02} & \revision{\textbf{91.95}} & 65.30 & \revision{56.48} & 83.11 & \revision{92.12} & \textbf{89.95} & \revision{\textbf{94.88}} & 3.1 & 20.7 & 33.3 \\
\textbf{Harbor} & 283 & 53.71 & 34.98 & \revision{47.63} & 26.86 & \revision{83.15} & \textbf{51.59} & \revision{\textbf{91.44}} & 34.98 & \revision{58.84} & 36.04 & \revision{72.43} & \textbf{54.42} & \revision{\textbf{97.95}} & 2.3 & 23.7 & 27.5 \\
\midrule
\textit{Avg / Total} & 1186 & 56.24 & 54.72 & \revision{51.70} & 46.54 & \revision{89.07} & \textbf{68.04} & \revision{\textbf{90.75}} & 51.43 & \revision{50.27} & 63.66 & \revision{80.83} & \textbf{72.18} & \revision{\textbf{93.37}} & 14.6 & 129.1 & 196.4 \\
\bottomrule
\end{tabular}%
}
\end{table*}

\begin{table*}[t]
\centering
\caption{\textbf{RQ2: Efficiency Analysis.} Error distribution (W=Wildcard, C=Condition Missing, NP=Not Predicted)\revision{, token consumption (In=Input, Out=Output, in K tokens), and average tool-call count per entry point (CC)}. Sta. = Static Reachability.}
\label{tab:rq2}

\small

\begin{tabular}{l c c c c c c c c c c c c c c c c c}
\toprule
\multirow{2}{*}{\textbf{Repository}} & \textbf{Sta.} & \multicolumn{5}{c}{\textbf{iCodeReviewer}} & \multicolumn{5}{c}{\textbf{RepoAudit}} & \multicolumn{5}{c}{\textbf{AutoSQL}} \\
\cmidrule(lr){2-2} \cmidrule(lr){3-7} \cmidrule(lr){8-12} \cmidrule(l){13-17}
& NP & W & C & NP & In/Out & \revision{CC} & W & C & NP & In/Out & \revision{CC} & W & C & NP & In/Out & \revision{CC} \\
\midrule
\multicolumn{17}{c}{\textit{Kimi}} \\
\midrule
\textbf{Answer} & 66 & 9 & 87 & 1 & 1.7/1.5 & \revision{0} & 9 & 82 & 17 & 22.5/0.6 & \revision{5.11} & 2 & 47 & 7 & 75.6/1.1 & \revision{8.29} \\
\textbf{Gitea} & 232 & 18 & 115 & 12 & 2.2/1.3 & \revision{0} & 52 & 135 & 9 & 13.9/0.5 & \revision{3.00} & 18 & 111 & 1 & 84.3/1.2 & \revision{8.22} \\
\textbf{Go-Admin} & 59 & 11 & 30 & 8 & 1.8/1.0 & \revision{0} & 26 & 28 & 0 & 58.4/1.0 & \revision{7.17} & 0 & 21 & 0 & 264.1/2.0 & \revision{19.13} \\
\textbf{Grafana} & 31 & 13 & 48 & 1 & 2.4/1.4 & \revision{0} & 17 & 43 & 9 & 11.1/0.6 & \revision{2.88} & 6 & 28 & 1 & 58.5/1.2 & \revision{6.98} \\
\textbf{Harbor} & 131 & 3 & 178 & 3 & 2.1/1.7 & \revision{0} & 5 & 195 & 7 & 24.3/0.7 & \revision{4.95} & 4 & 130 & 3 & 88.2/1.3 & \revision{9.03} \\
\midrule
\textbf{Total / Avg.} & 519 & 54 & 458 & 25 & 2.1/1.4 & \revision{0} & 109 & 483 & 42 & 19.7/0.6 & \revision{3.89} & 30 & 337 & 12 & 94.0/1.2 & \revision{9.02} \\
\midrule
\multicolumn{17}{c}{\textit{Claude Sonnet}} \\
\midrule
\textbf{Answer} & 66 & 4 & 74 & 0 & 1.6/1.2 & \revision{0} & 2 & 49 & 5 & 74.8/3.0 & \revision{6.57} & 2 & 39 & 0 & 95.9/1.8 & \revision{8.96} \\
\textbf{Gitea} & 232 & 18 & 160 & 10 & 2.4/0.9 & \revision{0} & 23 & 106 & 2 & 79.2/2.3 & \revision{5.49} & 26 & 87 & 1 & 143.4/2.0 & \revision{10.71} \\
\textbf{Go-Admin} & 59 & 10 & 32 & 8 & 2.1/0.7 & \revision{0} & 12 & 8 & 6 & 264.2/2.6 & \revision{14.83} & 9 & 15 & 0 & 262.7/2.6 & \revision{19.13} \\
\textbf{Grafana} & 31 & 23 & 52 & 1 & 2.9/0.9 & \revision{0} & 18 & 19 & 0 & 43.9/1.8 & \revision{4.18} & 7 & 14 & 1 & 78.0/1.9 & \revision{7.67} \\
\textbf{Harbor} & 131 & 3 & 177 & 4 & 1.5/0.9 & \revision{0} & 3 & 175 & 3 & 65.5/3.8 & \revision{6.10} & 7 & 121 & 1 & 143.3/1.9 & \revision{13.27} \\
\midrule
\textbf{Total / Avg.} & 519 & 58 & 495 & 23 & 2.2/0.9 & \revision{0} & 58 & 357 & 16 & 85.3/2.6 & \revision{6.25} & 51 & 276 & 3 & 136.4/2.0 & \revision{11.09} \\
\bottomrule
\end{tabular}%
\end{table*}

\subsection{RQ1: Effectiveness}

Table~\ref{tab:rq1} presents the recall, prediction \revision{precision}, and processing time (excluding code index construction) comparison between \alias and existing SQL auditing approaches across five subject systems. Since Claude Sonnet 4.5 is accessed through an external API with variable network latency and rate limiting, we only report processing time for Kimi.

\textbf{Dynamic features limit static reachability at 56.24\% recall.} The remaining 43.76\% of statements rely on reflection-based hooks, runtime schema resolution, and stateful builders, demonstrating that effective auditing must actively resolve dynamic runtime behaviors rather than relying solely on syntactic extraction.

\textbf{\alias addresses baseline failure modes through hybrid retrieval and ORM-aware synthesis.} Unlike iCodeReviewer, whose one-shot collection leaves \textit{Condition Missing} as the dominant error, and RepoAudit, whose rigid graph traversal misses non-structural dependencies and leads to \textit{Not Predicted} errors, \alias achieves 68.04\% \textasciitilde~72.18\% recall. It mitigates these issues via Hybrid Context Retrieval, combining the Code Index with \textit{ad-hoc} pattern-based search to escape structural dead ends. Additionally, its ORM Implicit Behavior Analysis identifies framework-specific semantics from struct tags and lifecycle hooks, resolving dynamic columns and operations to reduce \textit{Wildcard} errors. Consequently, \alias achieves \revision{an estimated precision of 90.75\%/93.37\% (Kimi/Claude). Compared with RepoAudit (89.07\%/80.83\%), AutoSQL achieves substantially higher recall while maintaining comparable or higher estimated precision, suggesting that broader control-flow enumeration does not necessarily introduce additional false positives. iCodeReviewer's estimated precision (51.70\%/50.27\%) reflects unmatched templates from one-shot generation}.

\subsection{RQ2: Efficiency}

Table~\ref{tab:rq1} and Table~\ref{tab:rq2} present the effectiveness and efficiency comparison between \alias and existing SQL auditing approaches. 

\alias consumes 94.0K to 136.4K input tokens per entry point and takes 196.4 minutes in total, surpassing both baselines in resource usage. However, this overhead is justified by its higher recall. The token cost is not due to inefficient retrieval but stems from the complexity of SQL reconstruction, which requires assembling fragmented logic from across the codebase. For instance, its 44.8$\times$ to 62.0$\times$ higher token usage compared to iCodeReviewer directly yields a recall gain of 13.32\% to 20.75\%, demonstrating that a minimal context is insufficient. Compared to RepoAudit, the increased cost is attributable to our methodology, such as Hybrid Context Retrieval and implicit ORM analysis, which enable a more thorough control-flow enumeration. This comprehensive context is essential for reasoning about control flow and cannot be significantly compressed without sacrificing accuracy. \revision{In terms of agent iterations, \alias averages 9.02/11.09 tool calls per entry point (Kimi/Claude), compared to 3.89/6.25 for RepoAudit; the increase reflects the additional structural traversal and pattern-based search needed to resolve cross-file dependencies.}

\subsection{RQ3: Ablation Study}


\revision{To quantify the contribution of our Hybrid Context Retrieval design, we compare AutoSQL against three representative agent paradigms, each employing a distinct retrieval or control strategy. Table~\ref{tab:rq3} presents the detailed ablation results.}

\begin{table*}[t]
\centering
\caption{\textbf{RQ3: Ablation Study.} (a) Recall (\%) and \revision{Estimated Precision (P, \%)}. (b) Error distribution (W=Wildcard, C=Condition Missing, NP=Not Predicted) and token consumption. Bash = Bash-Only, Vec = Vector, MA = Multi-Agent, AS = \alias.}
\label{tab:rq3}

\small
\textbf{(a) Recall \& \revision{Estimated Precision}}
\vspace{0.3em}

\resizebox{\textwidth}{!}{%
\begin{tabular}{l r c c c c c c c c c c c c c c c c}
\toprule
\multirow{3}{*}{\textbf{Repo}} & \multirow{3}{*}{\textbf{GT}} & \multicolumn{8}{c}{\textbf{Kimi K2}} & \multicolumn{8}{c}{\textbf{Claude Sonnet 4.5}} \\
\cmidrule(lr){3-10} \cmidrule(l){11-18}
& & \multicolumn{2}{c}{\textbf{Bash}} & \multicolumn{2}{c}{\textbf{Vec}} & \multicolumn{2}{c}{\textbf{MA}} & \multicolumn{2}{c}{\textbf{AS}} & \multicolumn{2}{c}{\textbf{Bash}} & \multicolumn{2}{c}{\textbf{Vec}} & \multicolumn{2}{c}{\textbf{MA}} & \multicolumn{2}{c}{\textbf{AS}} \\
\cmidrule(lr){3-4} \cmidrule(lr){5-6} \cmidrule(lr){7-8} \cmidrule(lr){9-10} \cmidrule(lr){11-12} \cmidrule(lr){13-14} \cmidrule(lr){15-16} \cmidrule(l){17-18}
& & R & \revision{P} & R & \revision{P} & R & \revision{P} & R & \revision{P} & R & \revision{P} & R & \revision{P} & R & \revision{P} & R & \revision{P} \\
\midrule
\textbf{Answer} & 145 & 47.59 & \revision{81.97} & 48.28 & \revision{85.29} & 52.41 & \revision{82.81} & \textbf{61.38} & \revision{\textbf{85.99}} & 33.79 & \revision{54.07} & 55.86 & \revision{81.20} & 44.83 & \revision{77.43} & \textbf{71.72} & \revision{\textbf{82.42}} \\
\textbf{Gitea} & 454 & 66.74 & \revision{87.63} & 65.64 & \revision{91.03} & 62.11 & \revision{\textbf{93.59}} & \textbf{71.37} & \revision{91.65} & 67.18 & \revision{87.17} & 68.50 & \revision{\textbf{96.88}} & 69.16 & \revision{68.76} & \textbf{74.89} & \revision{93.96} \\
\textbf{Go-Admin} & 85 & 69.41 & \revision{88.29} & 72.94 & \revision{81.54} & 74.12 & \revision{\textbf{94.84}} & \textbf{75.29} & \revision{86.58} & 69.41 & \revision{90.57} & \textbf{77.65} & \revision{86.36} & 70.59 & \revision{93.04} & 71.76 & \revision{\textbf{94.85}} \\
\textbf{Grafana} & 219 & 76.26 & \revision{88.57} & 73.97 & \revision{\textbf{94.21}} & 74.89 & \revision{91.26} & \textbf{84.02} & \revision{91.95} & 81.28 & \revision{97.10} & 79.45 & \revision{97.14} & 87.21 & \revision{\textbf{97.30}} & \textbf{89.95} & \revision{94.88} \\
\textbf{Harbor} & 283 & 48.06 & \revision{93.56} & 45.23 & \revision{91.04} & 47.70 & \revision{\textbf{95.14}} & \textbf{51.59} & \revision{91.44} & 49.47 & \revision{89.15} & 51.59 & \revision{91.91} & 40.28 & \revision{97.66} & \textbf{54.42} & \revision{\textbf{97.95}} \\
\midrule
\textit{Avg} & 1186 & 61.89 & \revision{88.09} & 60.71 & \revision{90.62} & 60.71 & \revision{\textbf{92.23}} & \textbf{68.04} & \revision{90.75} & 61.64 & \revision{86.48} & 65.60 & \revision{\textbf{94.10}} & 62.73 & \revision{79.56} & \textbf{72.18} & \revision{93.37} \\
\bottomrule
\end{tabular}%
}

\vspace{1em}

\textbf{(b) Error Distribution \& Token Consumption}

\resizebox{\textwidth}{!}{%
\begin{tabular}{l c c c c c c c c c c c c c c c c c c c c}
\toprule
\multirow{2}{*}{\textbf{Repository}} & \multicolumn{5}{c}{\textbf{Bash-Only}} & \multicolumn{5}{c}{\textbf{Vector}} & \multicolumn{5}{c}{\textbf{Multi-Agent}} & \multicolumn{5}{c}{\textbf{AutoSQL}} \\
\cmidrule(lr){2-6} \cmidrule(lr){7-11} \cmidrule(lr){12-16} \cmidrule(l){17-21}
& W & C & NP & In/Out & \revision{CC} & W & C & NP & In/Out & \revision{CC} & W & C & NP & In/Out & \revision{CC} & W & C & NP & In/Out & \revision{CC} \\
\midrule
\multicolumn{21}{c}{\textit{Kimi K2}} \\
\midrule
\textbf{Answer} & 4 & 71 & 1 & 46.9/1.1 & \revision{8.41} & 2 & 73 & 0 & 62.6/1.0 & \revision{8.64} & 2 & 66 & 1 & 54.1/2.7 & \revision{18.83} & 2 & 47 & 7 & 75.6/1.1 & \revision{8.29} \\
\textbf{Gitea} & 38 & 113 & 0 & 80.0/1.3 & \revision{11.42} & 35 & 121 & 0 & 89.2/1.2 & \revision{11.18} & 37 & 135 & 0 & 100.8/3.1 & \revision{25.87} & 18 & 111 & 1 & 84.3/1.2 & \revision{8.22} \\
\textbf{Go-Admin} & 1 & 24 & 1 & 138.8/1.6 & \revision{17.43} & 2 & 14 & 7 & 138.6/1.4 & \revision{16.00} & 1 & 15 & 6 & 108.6/3.3 & \revision{28.70} & 0 & 21 & 0 & 264.1/2.0 & \revision{19.13} \\
\textbf{Grafana} & 16 & 33 & 3 & 71.3/1.4 & \revision{10.87} & 21 & 36 & 0 & 68.8/1.2 & \revision{9.48} & 11 & 44 & 0 & 116.5/3.1 & \revision{24.38} & 6 & 28 & 1 & 58.5/1.2 & \revision{6.98} \\
\textbf{Harbor} & 3 & 142 & 2 & 74.7/1.3 & \revision{10.21} & 5 & 147 & 3 & 78.7/1.1 & \revision{9.85} & 5 & 142 & 1 & 102.8/3.7 & \revision{26.68} & 4 & 130 & 3 & 88.2/1.3 & \revision{9.03} \\
\midrule
\textbf{Total / Avg.} & 62 & 383 & 7 & 78.3/1.3 & \revision{11.24} & 65 & 391 & 10 & 84.8/1.2 & \revision{10.76} & 56 & 402 & 8 & 98.7/3.2 & \revision{25.13} & 30 & 337 & 12 & 94.0/1.2 & \revision{9.02} \\
\midrule
\multicolumn{21}{c}{\textit{Claude Sonnet 4.5}} \\
\midrule
\textbf{Answer} & 2 & 90 & 4 & 51.9/2.9 & \revision{7.33} & 2 & 61 & 1 & 86.2/1.7 & \revision{9.81} & 2 & 78 & 0 & 126.9/3.4 & \revision{18.56} & 2 & 39 & 0 & 95.9/1.8 & \revision{8.96} \\
\textbf{Gitea} & 35 & 107 & 7 & 118.4/2.0 & \revision{14.59} & 43 & 100 & 0 & 103.1/1.8 & \revision{11.90} & 39 & 100 & 1 & 249.7/5.7 & \revision{36.50} & 26 & 87 & 1 & 143.4/2.0 & \revision{10.71} \\
\textbf{Go-Admin} & 13 & 13 & 0 & 153.3/2.7 & \revision{17.04} & 2 & 9 & 8 & 170.5/2.1 & \revision{18.85} & 7 & 18 & 0 & 207.9/5.0 & \revision{30.17} & 9 & 15 & 0 & 262.7/2.6 & \revision{19.13} \\
\textbf{Grafana} & 20 & 21 & 0 & 78.1/1.8 & \revision{10.41} & 14 & 31 & 0 & 62.0/1.6 & \revision{8.24} & 5 & 23 & 0 & 113.1/3.2 & \revision{18.95} & 7 & 14 & 1 & 78.0/1.9 & \revision{7.67} \\
\textbf{Harbor} & 7 & 134 & 2 & 133.5/2.0 & \revision{13.24} & 8 & 126 & 3 & 115.2/1.7 & \revision{10.82} & 8 & 152 & 9 & 169.8/3.5 & \revision{19.05} & 7 & 121 & 1 & 143.3/1.9 & \revision{13.27} \\
\midrule
\textbf{Total / Avg.} & 77 & 365 & 13 & 109.0/2.2 & \revision{13.00} & 69 & 327 & 12 & 101.7/1.7 & \revision{11.42} & 61 & 371 & 10 & 195.9/4.6 & \revision{28.05} & 51 & 276 & 3 & 136.4/2.0 & \revision{11.09} \\
\bottomrule
\end{tabular}%
}
\end{table*}

\revision{\textbf{Hybrid Context Retrieval generally outperforms all three mainstream agent paradigms.}} Structural navigation via the Code Index shows consistent advantages over alternative retrieval strategies. \alias outperforms the Bash-Only agent by 2.35\% to 37.93\% across individual projects, confirming that purely pattern-based search may miss structural dependencies. Vector-RAG achieves similar recall to Bash-Only, indicating that embedding-based similarity struggles with structural code relations. Furthermore, \alias yields the lowest \textit{Condition Missing} errors (337/276 for Kimi/Claude) because structural graph traversal ensures more comprehensive context collection. In contrast, the Multi-Agent approach suffers more \textit{Condition Missing} errors, especially on Claude, as task decomposition divides the necessary global reasoning context across isolated agents. Across all methods, \revision{estimated precision remains generally high across all methods (88.09\% \textasciitilde~92.23\% for Kimi, 79.56\% \textasciitilde~94.10\% for Claude), indicating that false-positive rates are low regardless of retrieval paradigm. Methods with more thorough context retrieval tend to enumerate more control-flow paths, which can introduce a small number of additional false positives; however, this tradeoff is modest compared to the recall gains}. \revision{Since these three baselines collectively cover the core retrieval mechanisms (shell-based ReAct, embedding-based RAG) and control paradigms (hierarchical decomposition) of general-purpose coding agents, \alias's overall advantage suggests that Hybrid Context Retrieval, which combines structural graph navigation with on-demand pattern-based search, plays an important role in SQL reconstruction and is unlikely to be substituted by scaffold complexity alone.}

\textbf{Token consumption reveals the efficiency trade-offs among different retrieval paradigms.} \alias consumes more input tokens than the Bash-Only and Vector-RAG baselines (averaging 94.0K/136.4K vs. 78.3K \textasciitilde~84.8K / 101.7K \textasciitilde~109.0K for Kimi/Claude). This increase occurs because resolving cross-file dependencies, such as implicit hooks and type definitions, requires fetching additional code slices to mitigate \textit{Condition Missing} errors. Conversely, Vector-RAG retrieves code based on semantic similarity, introducing structurally irrelevant chunks with only mild improvements in recall. Furthermore, the Multi-Agent paradigm incurs the highest output token overhead (3.2K/4.6K vs. 1.2K/2.0K for \alias) due to inter-agent coordination. Consequently, \alias achieves higher recall by directing the token budget toward structural context rather than semantic similarity searches or inter-agent communication. \revision{This is also reflected in tool-call counts: Bash-Only averages 11.24/13.00 calls per entry point (Kimi/Claude) due to exploratory search, while \alias requires only 9.02/11.09 calls by leveraging Code Index navigation; the Multi-Agent paradigm incurs the most at 25.13/28.05, driven by inter-agent task delegation.}

\begin{table*}[t]
\centering
\caption{\textbf{RQ4: Parameter Sensitivity Analysis.} Impact of upstream tracing depth $D$ on recall and token consumption using Kimi K2.}
\label{tab:rq4}

\small

\begin{tabular}{l r r r r r r r r r r}
\toprule
\multirow{2}{*}{\textbf{Repository}} & \multirow{2}{*}{\textbf{GT}} & \multicolumn{2}{c}{\textbf{D=0}} & \multicolumn{2}{c}{\textbf{D=2}} & \multicolumn{2}{c}{\textbf{D=4}} & \multicolumn{3}{c}{\textbf{Token (In/Out, K)}} \\
\cmidrule(lr){3-4} \cmidrule(lr){5-6} \cmidrule(lr){7-8} \cmidrule(l){9-11}
& & Recall & Hits & Recall & Hits & Recall & Hits & D=0 & D=2 & D=4 \\
\midrule
\textbf{Answer} & 145 & 65.52 & 95 & 63.45 & 92 & 61.38 & 89 & 31.7/0.7 & 75.5/1.1 & 75.6/1.1 \\
\textbf{Gitea} & 454 & 52.64 & 239 & 75.55 & 343 & 71.37 & 324 & 64.6/1.1 & 96.2/1.2 & 84.3/1.2 \\
\textbf{Go-Admin} & 85 & 1.18 & 1 & 62.35 & 53 & 75.29 & 64 & 122.4/1.2 & 251.4/1.9 & 264.1/2.0 \\
\textbf{Grafana} & 219 & 84.93 & 186 & 88.13 & 193 & 84.02 & 184 & 55.6/1.1 & 75.9/1.3 & 58.5/1.2 \\
\textbf{Harbor} & 283 & 57.60 & 163 & 61.84 & 175 & 51.59 & 146 & 40.4/0.8 & 84.3/1.2 & 88.2/1.3 \\
\midrule
\textit{Average} & 1186 & 57.67 & 684 & 72.18 & 856 & 68.04 & 807 & 52.7/1.0 & 99.7/1.3 & 94.0/1.2 \\
\bottomrule
\end{tabular}%
\end{table*}

\subsection{RQ4: Parameter Sensitivity}

Table~\ref{tab:rq4} analyzes the impact of the upstream tracing depth parameter $D$ on recall and token consumption.

\textbf{The optimal tracing depth depends on project-specific ORM encapsulation.} Go-Admin requires $D=2$ for meaningful recall (jumping from 1.18\% to 62.35\%), indicating heavily encapsulated ORM calls. Conversely, Grafana achieves strong recall at $D=0$ (84.93\%) and peaks at $D=2$ (88.13\%); expanding $D$ further to 4 introduces noise, dropping recall to 84.02\%. Gitea improves from $D=0$ to $D=2$, while further increasing the depth to $D=4$ reduces recall. Token consumption mirrors these structural differences. For projects with shorter ORM invocation chains (e.g., Grafana), token usage actually decreases from $D=2$ to $D=4$. This happens because providing additional upstream context upfront prevents the agent from executing token-expensive supplementary searches during Phase III. In contrast, projects with very deep call chains (e.g., Harbor and Go-Admin) exhibit continuous token increases, with recall gains varying by project. Thus, optimal depth selection balances uncovering encapsulated logic against the noise and cost of excessive upstream layers.

\section{Threats to Validity}

\hspace{\parindent}
\textbf{Internal Validity.}
Our ground truth derives from runtime logs during test execution, covering only SQL statements triggered by tested paths. SQL statements in untested paths are absent from the ground truth, so a predicted SQL not in the logs may still be correct\revision{, preventing exact precision measurement. Our sampling-based estimation mitigates this but carries inherent sampling uncertainty}. Additionally, \alias emits wildcards when dependencies cannot be resolved through pattern-based search or when values are determined by external sources at runtime, and may miss SQL variants when complex control-flow conditions prevent complete path enumeration.

\textbf{External Validity.}
We evaluated our approach on five Go repositories (38K \textasciitilde~845K LoC) with \revision{two ORM frameworks (XORM and Beego), one custom SQL builder (Go-Admin), and three SQL dialects (SQLite, MySQL, PostgreSQL). Imperative ORM frameworks share a common high-level design such as method-chain builders, struct-tag column mappings, and lifecycle hooks, so AutoSQL's retrieval strategy transfers across frameworks. However, each ORM has its own terminal APIs and conventions; adapting to a new ORM requires updating the framework-specific knowledge in the system prompt. Our quantitative evaluation is scoped to test-covered entry points; results should not be generalized to untested entry points without further validation. As our selected open-source projects are well-known and likely present in LLM pre-training corpora, data contamination is a potential concern. However, our evaluation dataset is derived from instrumented test execution and expert annotation, which is unlikely to appear in any pre-training corpus. The models are thus evaluated on their ability to reconstruct these SQL statements from source code, not to recall memorized answers.}
\section{Related Work}

\subsection{SQL Synthesis from Code}

Static analysis techniques have been explored to synthesize or locate SQL queries from application code. Early work~\cite{QBS2012, QBS2013} applies program synthesis to convert imperative code into equivalent SQL queries. SLocator~\cite{SLocator} combines static analysis with information retrieval to locate SQL generation sites in Java applications. DBridge~\cite{Dbridge} constructs Java-to-database value flows through pointer analysis. These approaches are designed for Java's declarative ORM frameworks and cannot handle the stateful builder patterns and reflection-based mechanisms in Go ORMs. 

\subsection{LLM-Based Code Analysis and Auditing}

While LLMs have been applied to code analysis, recent evaluations~\cite{LLMcodeanalysis2024, PrimeVul2025} reveal they still struggle with real-world complexity and obfuscation. To mitigate this, various approaches decompose tasks for verification~\cite{LLMDFA2024, wang-etal-2024-sanitizing}, simulate pseudo-code execution~\cite{EVprompting2023}, or translate formal specifications~\cite{zheng-2025-validating}. For repository-level auditing, tools like iCodeReviewer~\cite{iCodeReviewer} and RepoAudit~\cite{RepoAudit} target vulnerability detection via single-pass or ReAct-style inference. Additionally, several methods integrate LLMs with static analysis: IRIS~\cite{IRIS2025} infers taint specifications, ErrorPrism~\cite{ErrorPrism2025} tracks error propagation, and RepoGraph~\cite{RepoGraph2025} leverages code graphs for structural navigation. However, these existing approaches primarily focus on security vulnerabilities or general code comprehension, lacking the tailored mechanisms required for the semantic reconstruction of dynamic database operations from imperative ORM code.
\section{Conclusion}


\revision{In this paper, we propose \alias, an LLM-based agent for extracting SQL templates from imperative ORM code.
Imperative ORMs construct SQL through dynamic method-call sequences, with many statements depending on runtime-resolved behaviors that are not explicitly represented in source-level structure.
Recovering these statements therefore requires non-structural context beyond direct code dependencies, which existing methods do not consistently capture.
To address this limitation, \alias combines structural code navigation with on-demand, pattern-based search to resolve dynamic dependencies and reconstruct SQL templates across these execution patterns.
Evaluation on a runtime-traced benchmark with 579 test-covered entry points from five large-scale repositories demonstrates that \alias achieves 68.04\%--72.18\% recall, surpassing static reachability by up to 15.94 percentage points and existing code audit agents by 8.52--21.50 percentage points.}

\begin{acks}
This work was supported by the National Natural Science Foundation of China (No. 62402536).
\end{acks}

\section*{Data Availability}

Artifacts and data: \url{https://doi.org/10.5281/zenodo.21872535}.

\balance
\bibliographystyle{ACM-Reference-Format}
\bibliography{sample-base}

\end{document}